\documentstyle[11pt,newpasp,twoside,epsf]{article} 
\begin{document}
\title{Detection limits in space-based transit observations} 
\author{H.J.Deeg} 
\affil{Instituto de Astrof\'{\i}sica de Andaluc\'{\i}a, E-18080 Granada, Spain; email hdeeg@iaa.es}
\affil{Instituto de Astrof\'{\i}sica de Canarias, E-38200 La Laguna, Tenerife. Spain}

\author{F. Favata and the \emph{Eddington} Science Team}
\affil{Astrophysics Division -- Space Science Department of ESA, ESTEC,
  Postbus 299, NL-2200 AG Noordwijk, The Netherlands}

\begin{abstract}
From simulations of transit observations, it is found that the
detectability of extrasolar planets depends only on two parameters:
The signal-to-noise ratio during a transit, and the number of data
points observed during transits.  All other physical parameters
describing transit configurations (planet and star size, orbital
period, orbital half axis, latitude of the transit across star) can be
reduced to these two parameters.  In turn, once the dependency between
transit detectability and these two parameters has been determined,
predictions for an instrument's ability to detect transits by any
combination of physical parameters can be derived with ease.  These
predictions are applied to the \emph{Eddington} proposal of a combined
Astroseismology/Transit-detection space mission currently under study
by ESA, which is described briefly.
\end{abstract}

\section{Introduction}
To evaluate the performance for the detection of extrasolar planets of
observations with the transit method, Jenkins, Doyle \& Culler (1996)
introduced the method of the calculation of two distributions, one of
the detection statistics for transit-less noise-only data (which may
be observed or simulated), and one detection statistics for data with
known transits in them. This method allows the determination of
detection probabilities while at the same time setting limits for the
false alarm rate - important in the case of transit detections, as one
neither knows if a transit signal is in the data, nor the signal's
duration, amplitude or period.  Doyle et al. (2000) used this method
to set detection limits in ground based observational data, taken with
the goal to detect transits across the eclipsing binary CM
Draconis. The evaluations shown here were performed during the course
of specifying the \emph{Eddington} proposal currently pending at ESA
for a combined astroseismology/planet-finding mission with a 1m-class
space telescope, which is described at the end.  In the following, the
dependence of the parameters describing a transit configuration
(Planet's size, star's size, planet's period for example) on
detectability is shown.

\section{Extrasolar planet detectability by transits}
Planet detection performance may be expressed in terms of
\emph{detectability} or the \emph{detection probability} $p_d$, which
is the probability that a physically present \emph{transiting} planet
is going to be detected by the experiment.  The detectability of a
planet in an observed transit sequence depends on a number of physical
parameters of the planet-star system and of the observing instrument.
The most obvious one is the size of the planet, with the amount of
brightness-loss during a transit being proportional to the square of
the planet radius, $R_{pl}$.  Similar, detectability will remain
roughly constant, if the ratio $R_{pl}/R_{*}$ is kept constant.

To asses the performance in the detection of planetary transits,
simulated lightcurves with the instrument's noise characteristics were
submitted to a \emph{detection test}.  This detection test gives some
'detection value', which may be calculated by any kind of procedure
that performs some evaluation between a given model lightcurve and the
data. The most suitable one for data which are dominated by `white'
noises and without (or with correctable) zero-point drifts is a scalar
multiplication between the data vector and the model vector:

                        \[C= \sum{d_{i}*m_{i}}\]

where $C$ is the detection value, and $d_{i}$, $m_{i}$ are data and
model values.  The detection test was performed in large numbers on
lightcurves with -- and without -- simulated transits added to the noise
The highest 'detection values' from a large number (between
$10^{5}-10^{7}$) of lightcurves \emph{without} transits (that is, with
noise only) give a threshold for detection values that constitute
false alarms. The fraction of lightcurves \emph{with} transits (again
using a large number of randomly generated curves, but with known
transit parameters) that results in detection values \emph{higher}
then the false alarm threshold gives the detection probability.  This
procedure is similar to one used for an assessment of detection limits
in the ground based transit detection project 'TEP' (Doyle, et al. 2000).


During the course of evaluating a variety of transit configurations,
it became obvious that the physical parameters governing transit
lightcurves may be reduced to two parameters which solely govern the
detectability in any observed sequence of transits around single stars
(the case for binaries would be more complicated): the data's S/N
during a transit, and the total number of data points within transits,
$N_{tp}$.  Thus, if the function $p_d(S/N, N_{tp})$ can be
established, the detectability can be calculated for any combination
of physical parameters. The physical parameters influence S/N and
$N_{tp}$ as follows:
\begin{list}{-}{}
\item
The Signal S depends on planet's and star's sizes, the latitude of
transit, and on the star's limb-darkening (mostly for high-latitude
transits). For a precise calculation of model transit lightcurves (and
hence S), a code dealing correctly with the planet's ingress and
egress and with differential limb-darkening under the planet's disk
should be used, such as `utm' (Deeg, 1999).
\item
The noise N depends on the noises affecting the observing instrument's
performance (which has always a dependency on a star's apparent
brightness) and on intrinsic brightness fluctuations (stellar
activity, oscillations, variability)
\item
The number of on-transit data points, $N_{tp}$, depends on
exposure-time, the number of transits observed (=observational
coverage) and on the duration of a transit.  The duration in turn is
governed by the planet's orbital period, by the star's mass and size,
by the latitude of the transit, and to a small amount by the size of
the planet. It can be calculated by straightforward application of
basic geometry and Kepler's third law.
\end{list}

For detectability evaluations, a 'standard case' was chosen, which has
a detection probability ($p_d$) of 0.99. In Fig. 1, the sequence of 6
transits of the standard case is shown. That case comprised the
following setup:
\begin{list}{-}{}

\item
Physical parameters: K0 star (0.85 R$_{sol}$, 0.78 M$_{sol}$), orbited by a planet with 1.4 R$_{\mathrm{Earth}}$, 200d orbital period, transiting at a latitude of 45\deg across the star.

\item
Observational parameters: 3 years (1095 days) observing duration;
lightcurve with a noise of 450 ppm, or approximately 0.5mmag from 600
seconds exposure time. In the case of the \emph{Eddington} proposal,
this is expected for a star with 16 mag.

\end{list}
 As an example of parameter variations around the standard 
case, the dependency of $p_d$ on planet-size is shown:

\begin{center}
\begin{tabular}{|c|c|c|} \hline
$R_{pl}/R_{\mathrm{Earth}}$& S/N & $p_d$ \\ \hline
1.0& 0.25& 0.16\\
1.3& 0.42& 0.91\\
1.4 & 0.48& 0.99\\
$\ga$1.5 &$\ga0.55$& 1.00\\ \hline
\end{tabular}
\end{center}

\begin{figure}
\plottwo{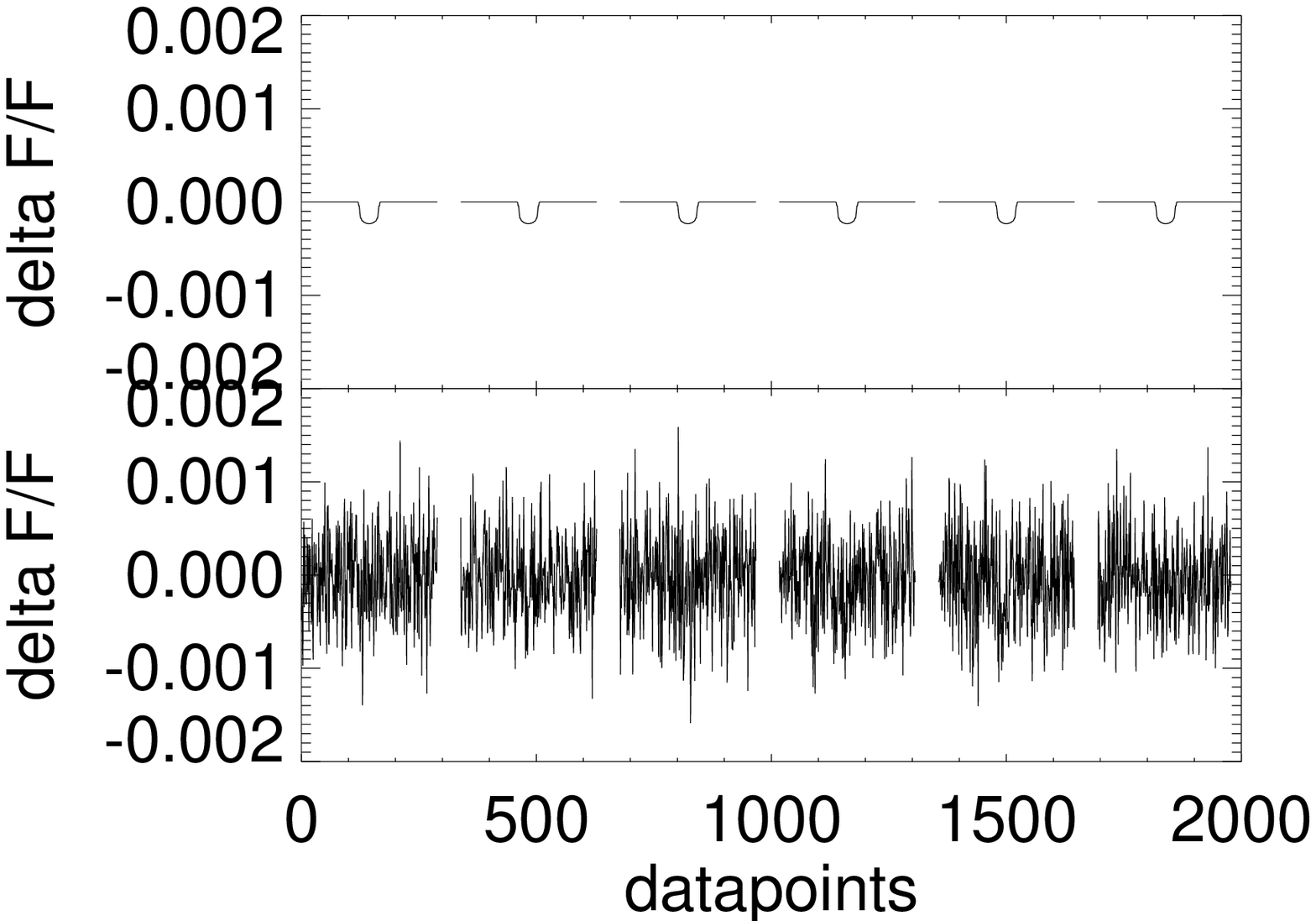}{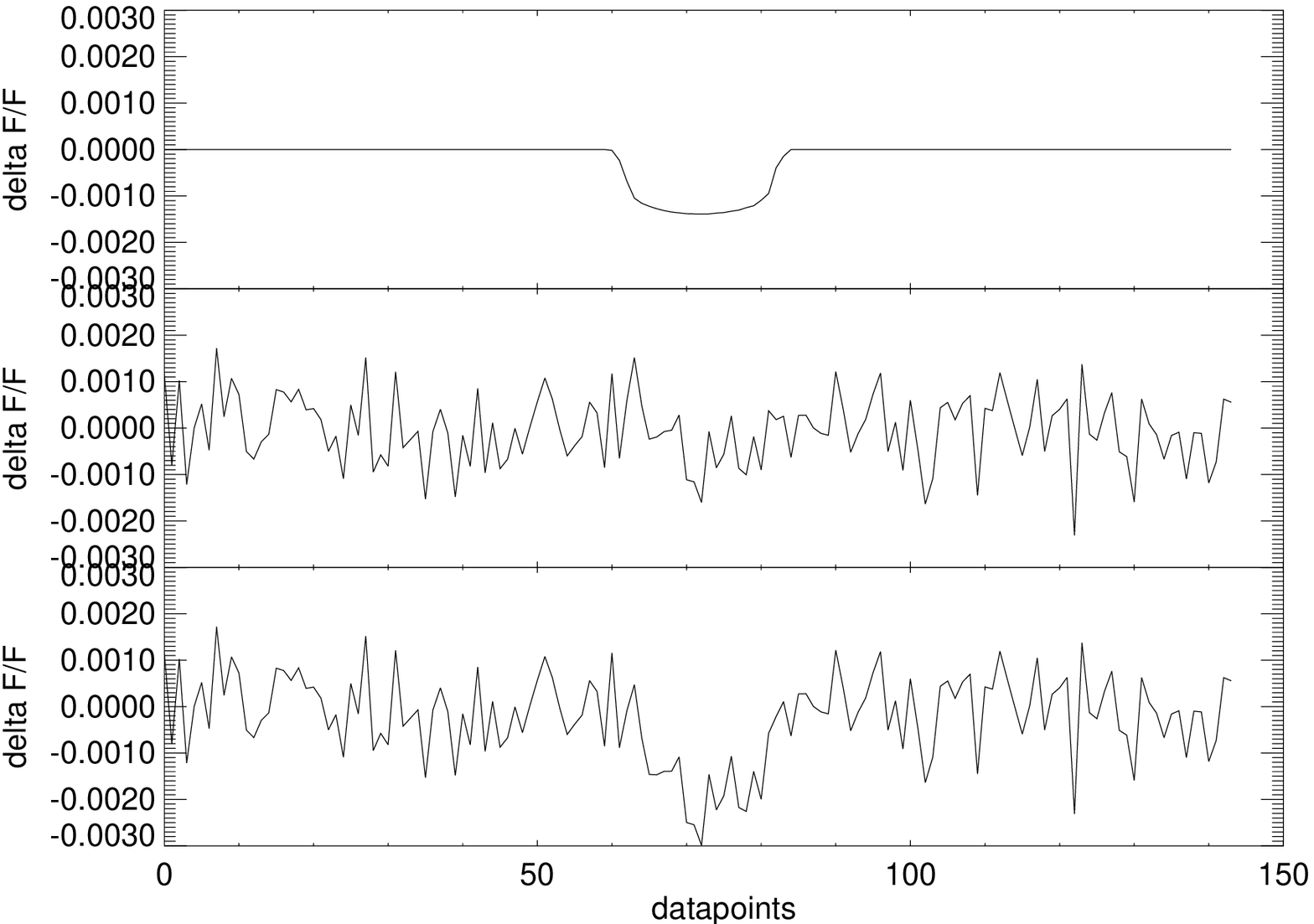}
\caption{Left: Sequence of 6 transits corresponding to observations of
the 'standard case'. The upper panel gives noiseless model transits,
the lower one is the simulated noisy lightcurve with the model
subtracted. Of the whole lightcurve, only segments near the 6 transits
are shown. As the detection limit of $p_d$=0.99 corresponds here to a
S/N of ~0.5, \emph{individual} transits are not detectable - neither to the
eye nor to any algorithm.  Right: the same transit sequence, but data
are rebinned by 2 and all 6 transits are coadded (that means: the
model-transit is 6 times deeper, and the noise is $\sqrt{6/2}$ times
larger). Center panel is pure noise, lower panel is
noise+transit. Planet detectability is practically independent of
exposure time or binning of the data. This holds, as long as noises
dominate that are proportional to $(\mathrm{exp.time})^{-1/2}$, such
as photon and background noise.}
\end{figure}

More useful then detection probability is however the determination of
the \emph{minimum planet size that can be detected reliably}. A
'reliable detection' is considered to have $p_d \ge 0.99$.  From the
table above, the detection limit is therefore a planet of 1.4
R$_{\mathrm{Earth}}$.  Similar tables where also generated for the
'standard case' but with orbital periods of 50, 100 and 365 days, thus
performing an empiric determination of the planet-size
$R_{\mathrm{pl,min}}$ where $p_d \approx 0.99$.  These results are
shown in the next table, where besides $R_{\mathrm{pl,min}}$ the
following quantities are given: $n_{tr}$, the number of transits in
the 3-year long lightcurve; $t_{tr}$, the duration of one transit;
$N_{tp}$, the total number of on-transit data points at 600 seconds
exposure time; and S/N, the signal to noise ratio, where the signal is
taken as the maximum relative brightness variation during a transit.

\begin{center}
\begin{tabular}{|l|c|c|c|c|} \hline
$P_{\mathrm{orb}}/\mathrm{day}$ & 50 & 100 & 200 & 365 \\ \hline
$R_{\mathrm{pl,min}}$&1.16&1.30&1.40&1.62\\
$n_{tr}$&22&11&6&3\\
$t_{tr}/day$&0.186&0.234&0.295&0.361\\
$N_{tp}$&589& 371 & 255 & 156\\
S/N& 0.33 & 0.42 & 0.48 & 0.64\\ \hline
\end{tabular}
\end{center}
  
From above table, a linear relationship between $\log S/N$ and $\log
N_{tp}$ was established, which can be expressed as:
\[ \log S/N = 0.90 -0.50 \log N_{tp}, \] 
or more generally: 

\[ S/N * \sqrt(N_{tp}) = \mathrm{const~} \mathrm{for~} p_{d} = \mathrm{const}. \]


It should be noted, that this expression is valid only when using
$C~=~\sum{d_{i}*m_{i}}$ for the determination of the `detection
value'.  Other possibilities to derive detection values, such as the
calculation of $\chi^2$ values between model and data, are expected to
lead to other dependencies, and may not allow the reduction of the
physical transit parameters to the two parameters S/N and $N_{tp}$.
Above expression between the fundamental parameters governing transit
detectability (using the level $p_d = 0.99$) allows the determination
of $R_{\mathrm{pl,min}}$ for any combination of physical parameters
constituting a transit configuration.


\begin{figure}
\plotone{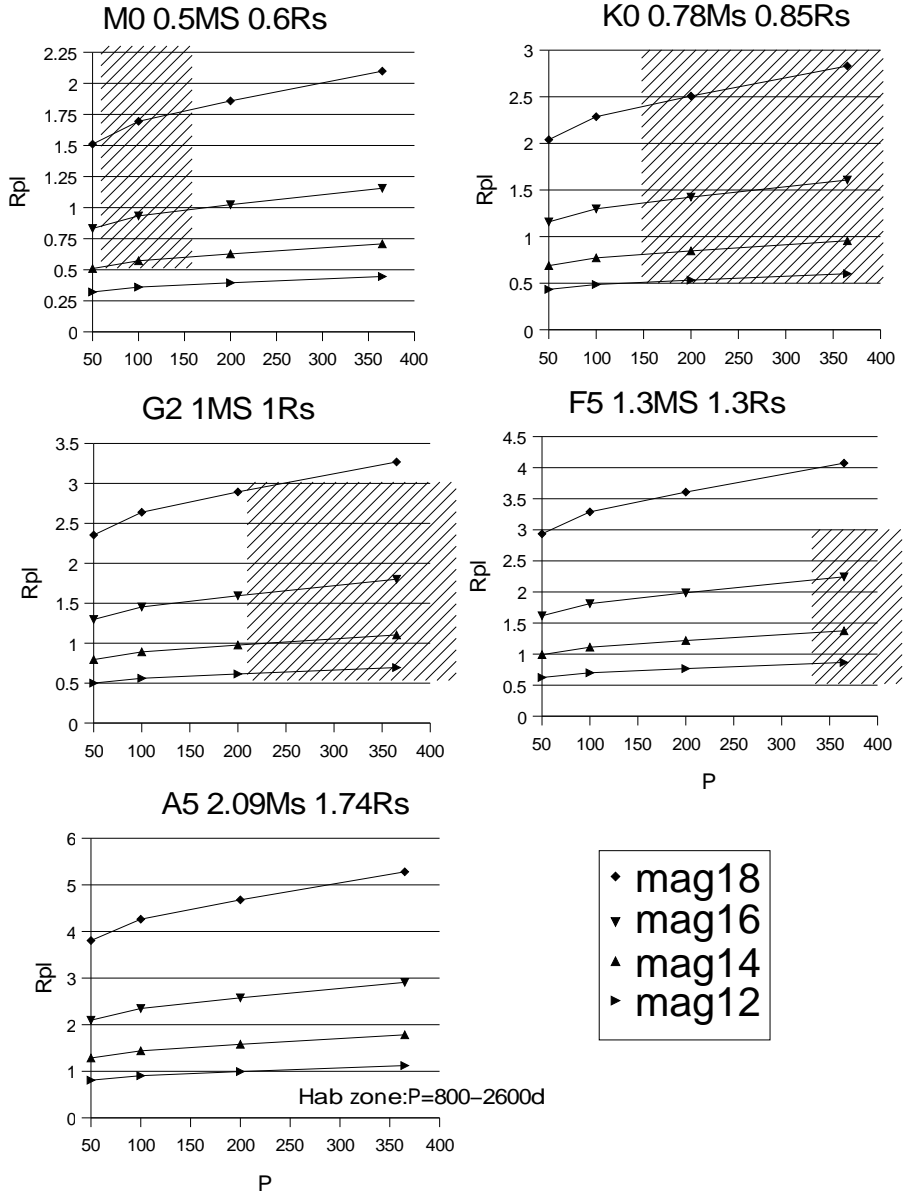}
\caption{Minimum detectable planet size (in $R_{\mathrm{Earth}}$) for a range of stars in
dependence of orbital period (in days) and apparent magnitude for the
\emph{Eddington} proposal. A transit latitude of 45\deg was
used. Hatched regions indicate the circumstellar habitable zone.}
\end{figure}

In Fig.~2, $R_{\mathrm{pl,min}}$ is given for a variety of stellar
types and orbital periods and apparent magnitudes as expected by
\emph{Eddington}. As the calculations were performed for transits at
latitudes of 45\deg, the given planet sizes are valid for $\approx$
70\% of theoretically possible transit configurations.  For the
remaining 30\% of polar transits, detection probabilities may be
significantly lower, and planets with $R_{\mathrm{pl,min}}$ cannot be
detected reliably.  Indicated as hatched regions in Fig. 2 is the
extent of the circumstellar habitable zone based on values from
Whitmire \& Reynolds (1996). It can be seen, that Earth-sized planets
around a solar-type star of $m_v$ = 14 may be detected, and even stars
with $m_v$ = 18 may still allow for the detection of 'large
terrestrial' planets.

\section {The \emph{Eddington} proposal}

This work was performed in the context of the \emph{Eddington} study.
\emph{Eddington} is a a combined asteroseismology and planet-detection
mission, based on a a 1~m class space telescope, initiated by a
European consortium of scientists (Favata, Roxburgh, \& Christensen-Dalsgaard 2000) in response to
ESA's F2/F3 mission call. \emph{Eddington} was successfully selected
for the initial currently ongoing study phase, and will undergo a
final selection in September 2000. With foreseen photometric precision
of $\simeq 1$ part in $100\,000$, the major science goal for the
planet-detection part will be the search for terrestrial size planets
in the habitable zone. With its large mirror size (1.2 m diameter) and
field of view (3\deg diameter), it will monitor several hundred
thousand stars during the 3 years of its mission dedicated to the
transit search. To enable long, uninterrupted observations within a
stable thermal environment, an L2 halo orbit is being baselined. These
parameters will allow \emph{Eddington} to achieve qualitatively
different science goals than the small missions currently in
preparation (COROT, MONS, MOST). The discovery of extrasolar
terrestrial planets, in addition to being a major achievement on its
own, will serve as a fundamental stepping stone for future missions
specifically designed for the detection of extraterrestrial habitats,
such as ESA and NASA missions IRSI-Darwin and Terrestrial Planet
Finder (TPF). \emph{Eddington} will also contribute in a major way to
the field of astroseismology, enabling an accurate characterization of
the stellar structure for stars spanning a wide range of masses, ages
and chemical compositions.

\end{document}